\documentclass[fleqn,twoside]{article}
\usepackage{espcrc2,natbib}

\usepackage{graphicx,epsfig}
\usepackage[figuresright]{rotating}

\newcommand{\gta}{\stackrel{>}{_{\sim}}}

\newcommand{\AmS}{{\protect\the\textfont2
  A\kern-.1667em\lower.5ex\hbox{M}\kern-.125emS}}

\newcommand{\aaps}{A\&AS}
\newcommand{\aap}{A\&A}
\newcommand{\apj}{ApJ}
\newcommand{\apjl}{ApJL}
\newcommand{\apjs}{ApJS}
\newcommand{\aj}{AJ}
\newcommand{\mnras}{MNRAS}
\newcommand{\nat}{Nature}
\begin{document}
\title{The accretion history of the Universe with the SKA}

\author{Matt J.~Jarvis\thanks{mjj@astro.ox.ac.uk} and Steve 
Rawlings\thanks{s.rawlings1@physics.ox.ac.uk}
\\{Astrophysics, Department of Physics, Keble Road, Oxford, OX1 3RH, UK }
}
\begin{abstract}
In this paper we investigate how the Square Kilometre Array (SKA) can
aid in determining the evolutionary history of active galactic nuclei
(AGN) from redshifts $z= 0 \rightarrow 6$.  Given the vast collecting
area of the SKA, it will be sensitive to both `radio-loud' AGN and the
much more abundant `radio-quiet' AGN, namely the radio-quiet quasars
and their `Type-II' counterparts, out to the highest redshifts. Not
only will the SKA detect these sources but it will also often be able
to measure their redshifts via the Hydrogen 21~cm line in emission
and/or absorption. We construct a complete radio luminosity function
(RLF) for AGN, combining the most recent determinations for powerful
radio sources with an estimate of the RLF for radio-quiet objects
using the hard X-ray luminosity function of \cite{Ueda03}, including
both Type-I and Type-II AGN.  We use this complete RLF to determine
the optimal design of the SKA for investigating the accretion history
of the Universe for which it is likely to be a uniquely powerful
instrument.

\vspace{1pc}
\end{abstract}

\maketitle

\section{INTRODUCTION}\label{sec:intro}

Radio astronomy has played an important r\^{o}le in the
hunt for high-redshift active galactic nuclei (AGN), with 
very-high-redshift
objects being pinpointed by radio pre-selection: at low radio frequency 
finding
`steep-spectrum' radio galaxies \citep{Raw96,vanB99} out to redshift $z 
\sim 5$; 
and at high frequency finding similarly distant
`flat-spectrum' (or `Giga-Hertz-Spectrum' GPS) quasars 
\citep[e.g.][]{Hook02}. 
The obvious benefits of radio pre-selection remain, e.g.\ the fact
that obscuration by dust and neutral Hydrogen is unimportant, contrasting 
dramatically with the case for optical and X-ray surveys. 

The reason why radio surveys led the way for so long is because a
radio-loud source is essentially detectable out to the very highest
redshifts with only a moderate exposure time, e.g.\ the highest
redshift radio galaxy discovered to date was selected from the
Westerbork Northern Sky Survey 
(WENSS\footnote{http://www.strw.leidenuniv.nl/wenss}; \cite{Reng97}) 
and has a flux density of $S_{365\rm MHz} \sim 0.6$~Jy. Another advantage of 
radio
surveys is that, particularly at the lower frequencies, the sky coverage 
per
pointing can be huge, e.g.\ for the new 74~MHz extension at the VLA
the field-of-view (FOV) is $\sim 40 ~ \rm deg^{2}$, much larger than
any FOV on an optical or X-ray telescope.

However, genuinely radio-loud sources are rare, probably $\sim
10 - 100$ times less abundant than their radio-quiet counterparts
\citep{Gold99}, whereas the optical \citep{Fan01a,Fan01b,Fan03}, and
(with the advent of XMM-Newton and CHANDRA) the X-ray surveys are now
taking the lead in surveying AGN in the high-redshift Universe
\citep[e.g.][]{Barger03a,Barger03b}. The results of these surveys are providing us with a much better understanding of both the
unobscured and obscured AGN populations, in the case of the Sloan
Digital Sky Survey \citep[SDSS;][]{sdssdr1} {\it and}, on the
traditional ground of the radio telescope, the X-ray satellites are
giving a much more transparent view of the more obscured population
\citep{Norman02}.  Unfortunately, there are still many unresolved
issues, most obviously highlighted by the fact that about 50\% of the
hard X-ray background is still unresolved into discrete AGN, the
consensus being that the current generation of X-ray telescopes,
operating at $0.5 \rightarrow 12$~keV, are missing the highly obscured
sources, with Hydrogen column densities $N_{\rm H} >
10^{28}$~m$^{-2}$ \citep{wilman00,Ueda03,Gandhi04}.

The SDSS and ultra-deep X-rays fields are paving the way 
\citep[e.g.][]{Brandt01,Alexander03b} 
but the quest to determine the nature and, possibly more
importantly, the evolution in the obscured population is liable to
switch back to the realm of radio astronomy with the commissioning of
the next generation of radio telescopes, most notably the
Low-Frequency Array (LOFAR\footnote{http://www.lofar.org}) and the
Square Kilometre Array (SKA\footnote{http://www.skatelescope.org}).

The LOFAR is currently set to operate at the very-low-frequency ($< 250 ~ \rm MHz$)
end of the radio spectrum, where the Universe has, to date, not been deeply
explored. This will allow an unprecedented view of the highest
redshift AGN, where the intrinsic spectral shape of a radio-loud AGN
means that the most sensitivity is gained at the lowest frequencies. 
Coupled with the fact that the 21~cm Hydrogen (HI) line is redshifted to
below 200~MHz at $z > 6$, where the epoch of reionization is coming to
an end \citep{Becker01}, this means that the LOFAR will be able
to probe the neutral Hydrogen within the epoch of reionization, as
opposed to the optical waveband where the Gunn-Peterson trough
due to neutral Hydrogen extinguishes Lyman-$\alpha$ photons
\citep{GP65}.  However the LOFAR , although important for
many studies of the AGN population, will not be able to secure
redshifts for the vast majority of the radio sources it finds, and as such 
optical and
near-infrared spectroscopy will always be needed. Moreover, due to the
large source density at the faintest flux-density limits, selection
criteria based on the radio spectral properties will still be required
in order to disentangle the highest redshift sources from the more
prominent low-redshift population 
\citep[e.g.][]{Huub94,Chambers96,Huub97,Blundell98,Jarvis01a,Jarvis01b,Carlos01}

\begin{figure}[ht]
{\hbox to \textwidth{\epsfxsize=0.48\textwidth \epsfbox{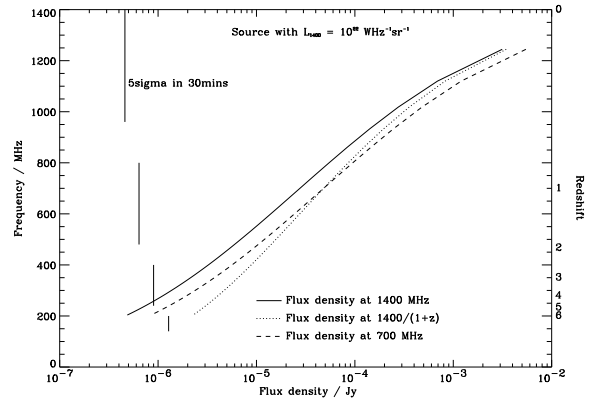} }}
{\caption{\label{fig:sens} The flux density of a `radio-quiet' AGN with $L_{1.4
      \rm GHz} = 10^{22}$~W~Hz$^{-1}$~sr$^{-1}$ and $\alpha = 0.8$, as a 
function of
    redshift and observing frequency. The solid line is the
    flux-density measured at the observed frequency of 1400~MHz; the
    dashed line is the flux density measured at 700~MHz and the dotted
    line is the flux density measured at the frequency of redshifted HI.}}
\end{figure}

A more promising telescope in terms of constraining the evolution and
properties of the AGN population is the SKA. With the proposed
sensitivities, the SKA will be able to detect all the
radio-loud and radio-quiet\footnotemark  AGN up to and including the epoch of 
reionization at $z >
6$ in a $\sim 30$-minute integration (Fig.~\ref{fig:sens}). The capabilities of the SKA for 
probing
AGN within the Epoch of Reionization are considered in Carilli et al.\ 
(this volume).
The SKA also has the capability to measure the
redshifts of many of the sources it detects via the HI
line in emission or absorption. A three-dimensional picture of the {\it 
whole} of the AGN
population will become apparent immediately.

\footnotetext{
A characteristic value of 
$L_{1.4 \rm GHz} = 10^{22}$~W~Hz$^{-1}$~sr$^{-1}$
is chosen for `radio-quiet' AGN above which (roughly speaking)
weak-AGN-dominated radio emitters dominate, and below
which starburst-dominated radio emitters dominate.
}

In this paper we investigate what the SKA will be able to contribute to 
studies of the whole of the AGN population, both `radio loud' and `radio quiet'
(see also Jackson, this volume; Falcke, K\"{o}rding \& Nagar, this volume).
In Section~\ref{sec:rlf} we describe the form of the radio luminosity
function (RLF) for the powerful radio-loud AGN. In
Section~\ref{sec:xlf} we show how to translate the hard-X-ray
luminosity function to that of an RLF for the
radio-quiet AGN population. Adding these together yields a {\it complete} 
radio luminosity
function for AGN, described in Section~\ref{sec:completerlf}. In section~\ref{sec:see} we estimate what the SKA will see in terms of AGN and in Section~\ref{sec:skadesign} we investigate the optimum 
design
for the SKA in terms of constraining the evolution of the entire AGN 
population.
In Section~\ref{sec:conclusions} we outline some of the science that the 
SKA is likely to deliver.

We adopt a $\Lambda$CDM cosmology with the following parameters:
$h = H_{0} / (100 ~ \rm km ~ s^{-1} ~ Mpc^{-1}) = 0.7$;
$\Omega_{\rm m} = 0.3$; $\Omega_{\Lambda} = 0.7$, throughout. We define 
radio spectral index 
as $S_{\nu} \propto \nu^{-\alpha}$, where $S_{\nu}$ is the radio flux 
density at frequency $\nu$.

\section{The radio-loud luminosity function}\label{sec:rlf}

The radio-loud population can be split into two types of source, with
a division (roughly speaking) at a critical radio luminosity in radio
structure \citep{FR74}, narrow-emission-line strength \citep{HL} and
quasar fraction \citep{CJWqsofrac}.  Below a radio luminosity
$L_{1.4\rm GHz} \sim 10^{25}$~W~Hz$^{-1}$~sr$^{-1}$, objects are
typically FRI radio galaxies with weak or absent emission lines. Above
the break in the RLF (at $L_{1.4\rm GHz} \sim
10^{26}$~W~Hz$^{-1}$~sr$^{-1}$), objects are a roughly equal mix of
FRII radio galaxies and FRII radio quasars, typically with strong
emission lines, indicating buried quasar nuclei even in the cases
where no direct quasar emission is evident. The transition region
between these two sub-populations lies just below the break in the
RLF.  The vast majority of work in constraining the evolution in the
luminosity function of the powerful radio sources has concentrated on
the `FRII sub-population', due mainly to the fact that the low radio
luminosities of FRI sources makes them hard to detect out to the
highest redshifts.

\subsection{The high-luminosity RLF: FRIIs}

The form of the RLF for the FRII sub-population has been constrained
to a certain degree by the work of \cite{DP90}, and more recently
\cite{CJWRLF}.  Here we use one of the parameterized models of Willott
et al.\ (model C) which describes the high-luminosity population by a
`reversed Schechter function' with an evolutionary term fitted with
two one-sided Gaussian functions, forced to match at a peak redshift
of $z = 1.9$. The form of this RLF means that the increase in source
density towards this redshift is much steeper than the higher redshift
decline. However, the data used by \cite{CJWRLF} to model the RLF has
a dearth of sources at high redshift due to a high flux-density limit
for the large area samples \citep[e.g. 3CRR;][]{LRL83} and a lack of
sky area for the faint flux-density samples \citep[6CE and
7CRS;][]{Eales97,SR01,7Cspec,7CKz}. Thus, the high-redshift evolution
is not yet well constrained.

The form of the decline in the comoving space density at high redshift
redshift has come under close scrutiny from a number of groups, some
of which suggest quite a steep decline at $z > 3$ \citep{Shaver96},
and some of which suggest a more moderate decline
\citep[e.g.][]{DP90}. However \cite{JR00} and \cite{Jarvis01c}
highlighted how uncertain the form of this decline is for both
flat-spectrum radio-loud quasars (which, crudely speaking, should
reflect the overall evolution as it picks out favourably oriented
sources from the underlying population) and also
low-frequency-selected radio galaxies, due predominantly to
difficulties in modelling the $k-$corrections, and the lack of
available volume (and hence small numbers of sources) at high redshift
for a given flux-density-limited sample.  This uncertainty is also
mirrored by the X-ray luminosity function
\citep[e.g.][see also Section~\ref{sec:xlf}]{Miyaji00,Ueda03}.  Our
adopted model lies between the extremes of a sharp cut off and an RLF
which stays constant with redshift at $z > 2$.  Such an evolution is
consistent with the predictions of simple Press-Schechter-based
theoretical models \citep{RJ04}.

\subsection{The moderate-luminosity RLF: FRIs}

Unfortunately, the RLF for the lower-luminosity radio-loud AGN,
predominantly (but not exclusively) comprised of FRI radio sources,
hereafter the `FRI sub-population', is poorly constrained. Several
groups have reached different conclusions concerning their
evolutionary behaviour \citep[e.g.][]{Jackson99,Snellen01,CJ04}. Here we adapt the results of
\citet{CJ04}, who used the largest dataset of $\sim 1000$ sources,
with a radio flux-density limit of $S_{325\rm MHz} = 40$~mJy, with
photometric redshifts determined by five-colour photometry of the
SDSS, to show that radio sources with $10^{23} \leq (L_{325\rm MHZ}$ /
W~Hz$^{-1}$~sr$^{-1}) \leq 10^{25}$ are consistent with having a
constant comoving density, at least in the redshift range $z = 0
\rightarrow 0.5$.  We also note that the results of their analysis are
consistent with that of \cite{Wadd01}, who constrained the form of the
evolution using a much fainter flux-density limited survey
\citep[LBDS;][]{Wadd00} but with a higher fraction of redshift
incompleteness. We allow this sub-population to have a comoving space
density which evolves very mildly with redshift as $\Phi \propto (1+z)^{1.2}$,
normalised to the comoving space density of this sub-population from
the model of Willott et al. at $z = 0.5$, i.e. where redshift data are
able to directly constrain the space density.

\section{The low-luminosity RLF: radio-quiets}\label{sec:xlf}

As stated in Section~\ref{sec:intro} the SKA will have the sensitivity
to trace radio-quiet quasars and radio-quiet type-II AGN out to
the highest redshifts. Therefore, we have attempted to
parameterize the form of the radio-quiet RLF and its evolution with
redshift. The work of \cite{Kukula98} suggested that the radio emission
from radio-quiet quasars is typically compact and is probably directly
associated with the central engine. They also showed that radio-quiet
quasars have spectral indices consistent with their more radio-loud
counterparts, i.e.\ $\alpha \sim 0.7$.

We also know that the X-ray luminosity of an AGN is a good tracer of
the accretion rate onto the SMBH. Thus, there should be a link between
the X-ray luminosity and the radio luminosity if both are regulated by
the accretion process. Indeed, \cite{Brinkmann00} show that this link
does exist. They compiled a sample of X-ray selected AGN from the
ROSAT All-Sky Survey \citep{Voges99} and cross-correlated this sample
with the 1.4~GHz FIRST survey \citep{White97}. They find a relatively
tight linear correlation between the logarithm of radio luminosity at
1.4~GHz and the X-ray luminosity in the soft X-ray band at 0.5-2~keV
band. The form of this correlation is

\begin{equation}
\label{eq:xray}
\log (L_{X}) = -4.57 + 1.012 \, \log (L_{1.4\rm GHz}),
\end{equation}

\noindent i.e.\ roughly a proportionality. With this in mind, it is
possible to use a complete description of the radio-quiet quasar
population in the X-ray band to make a {\it crude estimate} of the RLF
of radio-quiet objects at 1.4~GHz.

With the advent of XMM-Newton and Chandra, and the deep fields
surveyed with these satellites, a hard X-ray luminosity function (XLF)
has been determined by \cite{Ueda03}, building on the work of
\cite{Miyaji00} in the softer X-ray bands.  They find that the best
parameterisation of the XLF is one of luminosity-dependent density
evolution, i.e. one in which the cosmic evolution in the space density
of the sources depends on the intrinsic luminosity of the source. They
conclude that the comoving space density of the more X-ray-luminous
AGN peaked at an earlier epoch than those at lower X-ray
luminosities. The evidence for this type of evolution is reinforced by
other studies \citep[e.g.][]{Steffen03}. They also find that the
higher luminosity sources have less intrinsic absorption, consistent
with a `receding-torus-like' model \citep[e.g.][]{Lawrence91} which
has been suggested as a necessary physical ingredient to explain the
properties of radio-selected samples
\citep[e.g.][]{Simpson98,CJWqsofrac,Grimes04}.

With this parameterization, we predict how many X-ray sources there
will be of a given luminosity in the (rest frame) $2 \rightarrow
10$~keV energy range, at a given redshift.  We use the rest-frame
ratio of the average number of photons in this band to that of the
monochromatic luminosity at 1~keV for an unobscured source, noting
that the quasars were predominantly unobscured in the study of
\citet{Brinkmann00}.  We then convert this soft-X-ray luminosity to
the 1.4~GHz radio luminosity using Equation~\ref{eq:xray}.  In
addition, we need to take some account of a population of
highly-obscured (Compton thick) AGN, which, whilst essentially absent
from the existing `hard' (2-10 keV) surveys, would be revealed by
still harder X-ray surveys \citep{WilmanFabian99} with the next
generation of X-ray satellites
(e.g. XEUS\footnote{http://www.rssd.esa.int/XEUS} and
Constellation-X\footnote{http://constellation.gsfc.nasa.gov}) or
radio surveys. To do this we multiply the comoving space density
derived from the XLF by a conservative factor of 1.5 to account for
the X-ray obscured sources, which brings the integrated luminosity
closer to that expected from the hard-X-ray background.

\section{The complete RLF and constraints from the source 
counts}\label{sec:completerlf}

\begin{figure*}[ht]
{\hbox to \textwidth{\epsfxsize=0.98\textwidth \epsfbox{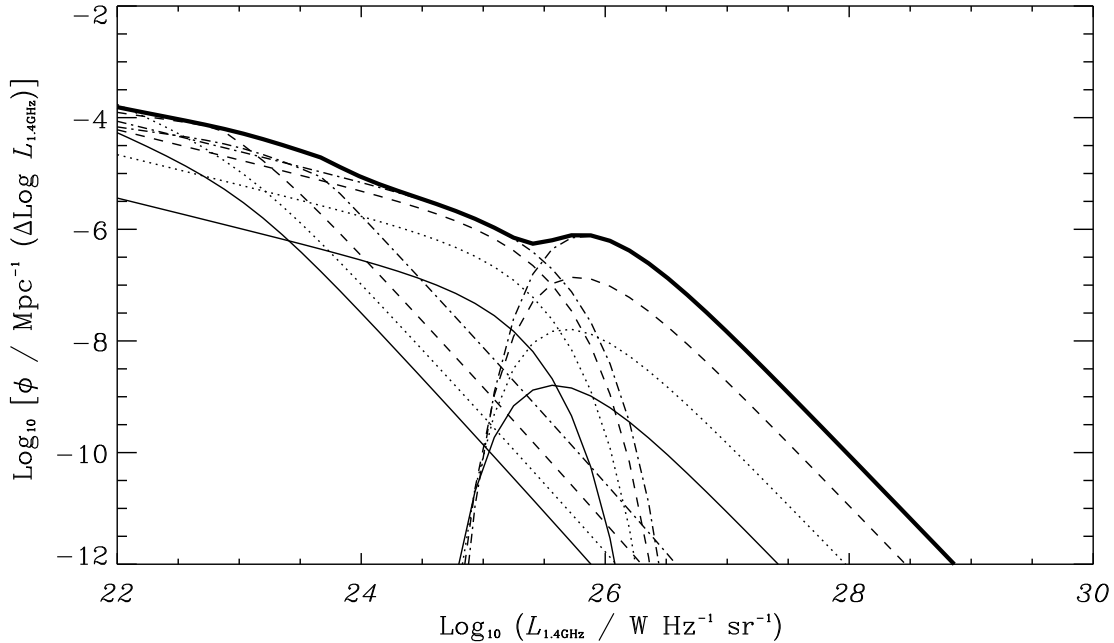} }}
{\caption{\label{fig:rlf} The 1.4~GHz radio luminosity function for
all AGN at four different redshifts. The contributions from the
radio-quiet quasar population can be seen to be significant 
between $10^{22} < (L_{1.4} / $W~Hz$^{-1}$~sr$^{-1} ) < 10^{24}$,
whereas the bulk of the radio luminosity density is provided by the
FRI-type sources at $L_{1.4} \sim
10^{25}$~W~Hz$^{-1}$~sr$^{-1}$. Finally, at $L_{1.4} >
10^{26}$~W~Hz$^{-1}$~sr$^{-1}$ all of the luminosity density is
provided by the powerful FRII-type sources. The thin solid line is the
RLF for each sub-population at $z = 0$, the dotted line is at $z =
0.5$, the dashed line is the RLF at $z = 1$ and the dot-dashed line is
the RLF at $z = 2$. To clarify the overall form, the thick solid line
represents the composite RLF, comprised of the FRII, FRI and
radio-quiet populations described in Section~\ref{sec:rlf} at $z = 2$.
}}
\end{figure*}

\begin{figure*}[ht]
{\hbox to \textwidth{\epsfxsize=0.98\textwidth \epsfbox{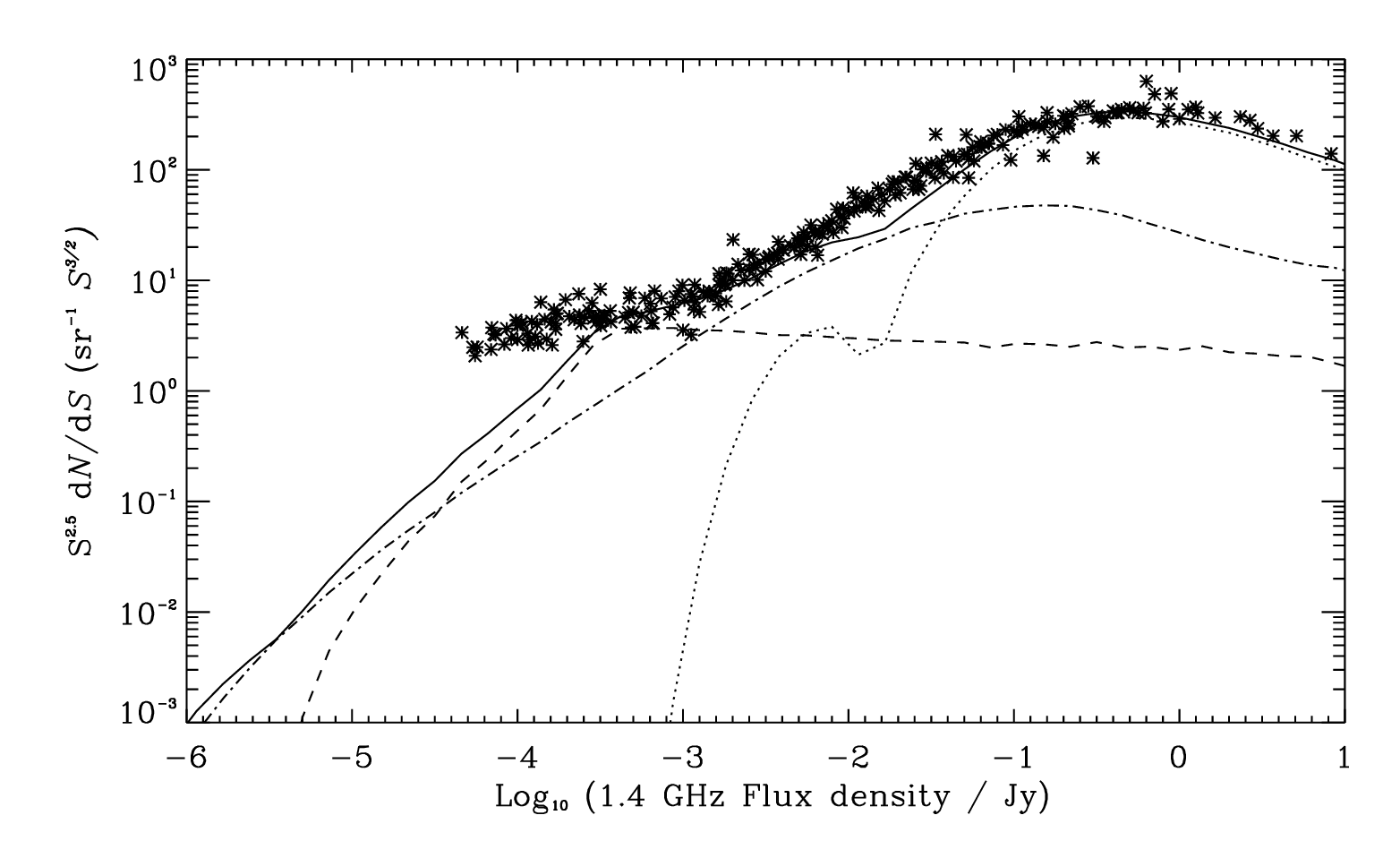} }}
{\caption{\label{fig:sourcecounts} The 1.4~GHz differential source
    counts (relative to counts in a Euclidean Universe)
of a combination of surveys \citep[see][for a description of 
the data points]{Seymour04}. The lines represent the
    model source counts from our three components of the radio
    luminosity function. The dotted line represents the powerful FRII
sources, the dot-dashed line is for the moderate luminosity, typically
FRI sources, and the dashed line represents the source counts for
radio-quiet quasars. The solid line is the total source counts from
these models.}}
\end{figure*}

In Figure~\ref{fig:rlf} we show the complete RLF, incorporating
the radio-loud sources (consisting of the two FRI and FRII
sub-populations following \citet{CJWRLF}) and the radio-quiet
quasars. This is the RLF we use to predict what the SKA will see in
terms of AGN for various SKA designs. With this parameterisation of the RLF we can make
a consistency check with the radio source counts at 1.4~GHz. In
Figure~\ref{fig:sourcecounts} we show the radio source counts at
1.4~GHz down to a flux-density limit of $S_{1.4 \rm GHz} \sim 1~\mu$Jy, along with the source counts derived from our RLF.

One can see immediately that the RLF traces the source counts
reasonably well at the high-flux-density regime, as it should do, as
this is where direct constraints from data exist. The small
deviations are artefacts of the `two-sub-population' method which
would be removed by using smoother functional forms (such as the
`free-form' fits of \citet{DP90} or the generalized luminosity
functions of \citet{Grimes04}. At lower radio flux densities the
moderate-luminosity `FRI' AGN begin to dominate the source counts, but
at lower flux densities still ($S_{1.4\rm GHz} \sim 5 \times 10^{-4} ~
\rm Jy$) the radio-quiet population can start to contribute
significantly to the source counts. The upturn in the source counts
around this flux density is often attributed solely to a low-redshift,
star-forming population. However as can be seen from
Fig.~\ref{fig:sourcecounts} the radio-quiet population, comprising
radio-quiet quasars and their `Type-II' counterparts (including
Compton-thick sources) may contribute significantly to this 
upturn (see also Baugh et al., this volume). Of
course, starburst-driven radio emission may make an important
contribution to the radio flux densities of at least some radio quiet
quasars \citep[e.g.][]{Sopp92,Prouten04}, and `starburst' galaxies seem commonly to
harbour X-ray (AGN) nuclei \citep[e.g.][]{Alexander03a}, so the
distinction between these two populations is not particularly clear.
The starburst population is considered elsewhere (van der Hulst et al., this
volume), and it is clear from Fig.~\ref{fig:sourcecounts} that as
we approach the depths of even shallow SKA surveys (surveys limited at
$\sim \mu\rm Jy$ levels), a small fraction of the objects are
likely to have their radio emission powered by any sort of AGN
activity.

Our description of the radio luminosity function, although very crude,
allows us to at least estimate the contributions from the various
populations of radio AGN.  As such we can distinguish between those
with massive black holes but low accretion rates (FRIs) and those with
massive black holes and high accretion rates (the radio-quiet
quasars). However, it is worth noting that \cite{KMBSR01} have shown
that both classes can produce similar radio luminosities and radio
structures. Ideally, a complete model for the evolution of all AGN in
radio wavebands would allow for more physically driven processes, with
a more continuous level of accretion between the various classes (or
equivalently less of a dichotomy between `radio-loud' and
`radio-quiet' sources).

Also, we have not attempted to add prescriptions for
favourably-oriented, and hence Doppler-boosted (flat-spectrum) sources
and young (GPS) sources as we will detect every
AGN with the SKA and these sources do not make up the bulk of the
population when probing such low flux-density limits.

Again, more realistic models could allow for factors such as the time
evolution of radio luminosity (although we caution that such models
bring in lots of uncertain physical details such as evolutions in
environment, and the behaviour of jet power as a function of time).

However, regardless of all such details, with a crude composite RLF in
place we are able to make a crude prediction of the number of each
type of AGN the SKA will be able to detect as a function of flux
density and area.

\section{What will the SKA see?}\label{sec:see}

\subsection{The number of AGN detected by the SKA in 1 year}

\begin{figure}[ht]
{\hbox to \textwidth{\epsfxsize=0.48\textwidth \epsfbox{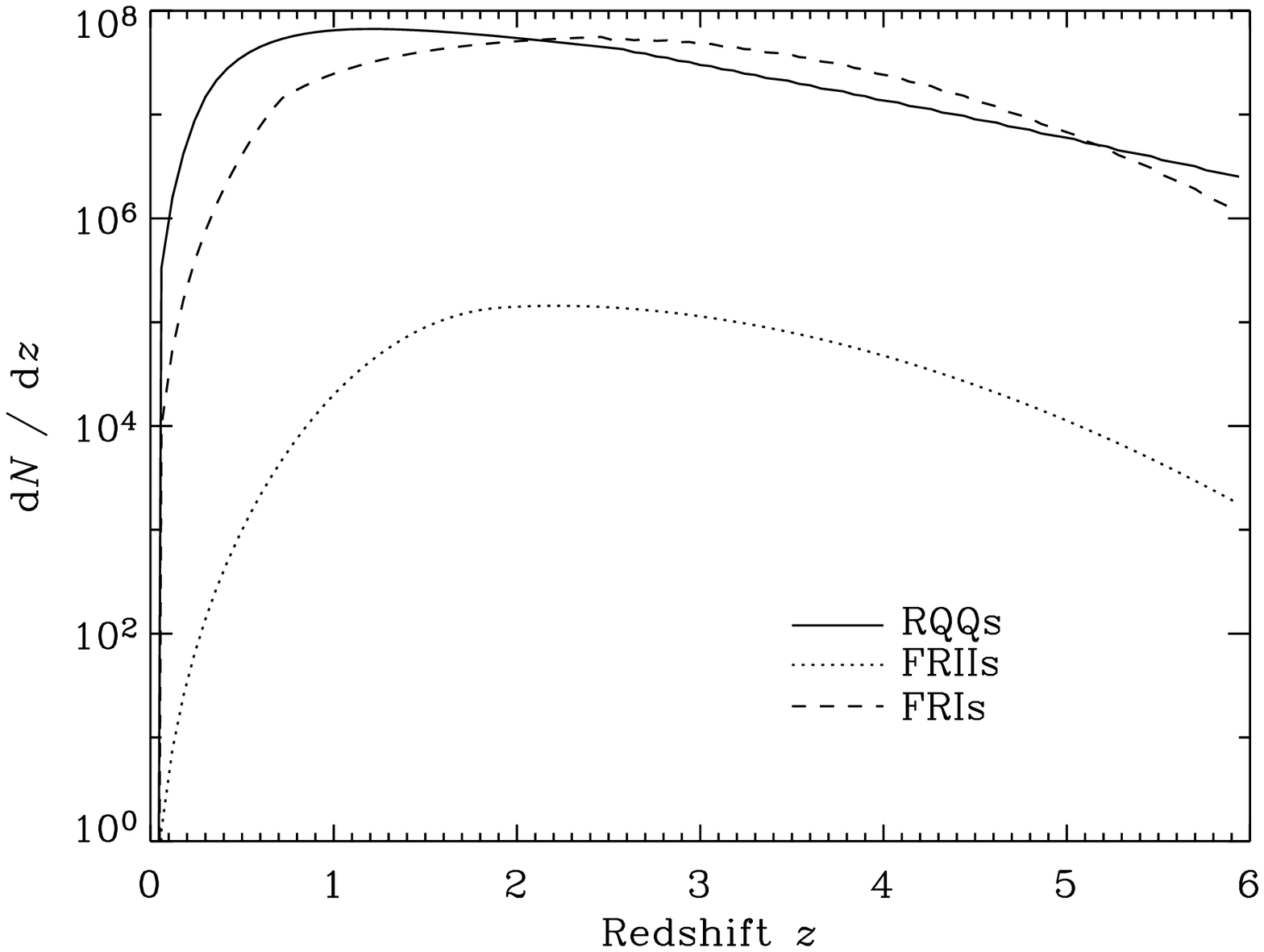}
}} {\caption{\label{fig:dndz} The number of the
different sub-classes of sources discussed in Section~\ref{sec:completerlf}
for a one year SKA survey over one hemisphere 
($\sim 20,000 ~ \rm deg^{2}$)
complete to $S_{200\rm MHz} \sim 100$~nJy. The dotted line
represents the powerful FRII-type sources, the dashed line shows the
evolution in the FRI-type sources and the solid line denotes the
radio-quiet quasars calculated using the XLF of \cite{Ueda03} and
multiplied by a factor of 1.5 (see Section~\ref{sec:xlf}).}}
\end{figure}

Using the composite RLF discussed in Section~\ref{sec:rlf} we are able
to estimate the number of sources that the SKA will be able to detect
as a function of both redshift and flux-density limit. In
Fig.~\ref{fig:dndz} we plot the predicted d$N/$d$z$ distribution for
each sub-population in an all-hemisphere SKA survey which would reach
a limiting ($4\sigma$) flux-density limit of $S_{200\rm MHz} = 100$~nJy
in one year of dedicated observations with the SKA\footnotemark.

\footnotetext{
This assumes a usable field of view at 200 MHz of $\approx 50 ~ \rm deg^{2}$
(or $FOV = 1 ~ \rm deg^{2}$ at 1.4 GHz in the diffraction-limited case), so
a survey of $\sim 20,000 ~ \rm deg^{2}$ requires $\sim 400$ pointings. A one-year 
survey (at 90 per cent observing efficiency), means $\approx$20 hours of 
exposure per pointing, yielding a survey rms of $\approx 25 ~ \rm nJy$. 
In practice, a `tiling' method will be used to ensure complete sky coverage
across the entire $0.2 \rightarrow 1.4 ~ \rm GHz$ range (see Rawlings et al., this
volume).   
For an SKA design with a $\sim$10-times-higher $FOV$, a 10-$\sigma$ catalogue 
to the same limiting flux density could be compiled from
a survey taking eight months.
}

One can see that the SKA will be able to detect every bona fide AGN in
the Universe up to $z \sim 6$. It is worth noting that there is a
large overlap between what we have called FRI-type AGN and the
radio-quiet quasars. This is because it is largely a question of
semantics as to what kind of source belongs where
\citep[e.g.][]{KMBSR01}.

It is also worth noting that at high redshift ($z > 3$) all three of
the sub-populations are largely unconstrained. Therefore, we force
both the FRI-type sources and the radio-quiet sources to decline
according to the decrease in space density expected under a
Press-Schechter based model \citep[see e.g.][]{RJ04} above a redshift
of $z \sim 3$. For the more powerful, FRII-type sources we just use
the shallow decline in the space density described by model C in
\cite{CJWRLF}.

Due to these large extrapolations the ratio between radio-loud sources
and radio-quiet sources stays roughly similar to what we observe at $z
< 2$ with today's telescopes. However, the SKA will actually measure this ratio to
great accuracy, and as a consequence may yield strong constraints
on AGN driven feedback events (see Section~\ref{sec:conclusions} and
\cite{RJ04}) and the duty cycle of all types of AGN.

\subsection{Supermassive black holes and their accretion rate}

In this section we estimate what the SKA will be able to see in
terms of the accretion rate of black holes with masses in the range
$10^{6}~{\rm M}_{\odot} < M_{\rm bh} < 10^{9}~{\rm M}_{\odot}$. To estimate the
bolometric luminosity, and thus calculate the fraction of the
Eddington luminosity which relates to a given radio luminosity for
both the powerful radio-loud sources and the radio-quiet sources we
use two separate methods.

For the radio-loud sources, i.e. those sources above the break in the
RLF (see Section~\ref{fig:rlf}) we use the relation of \cite{CJWemline}
which relates the low-frequency radio luminosity to the total ionizing
power of the AGN. This has the form,

\begin{equation}
\log(L_{\rm Bol} / \rm W) \approx 0.83 \log(L_{151} / {\rm W~Hz}^{-1}{\rm sr}^{-1}) + 16.9,
\end{equation}
where $L_{\rm Bol}$ is the bolometric luminosity and 
$L_{151}$ is the monochromatic luminosity at 151~MHz. We convert the 151~MHz luminosity to 200~MHz using a spectral index of $\alpha = 0.8$.

For the case of the radio-quiet quasars (and type-II obscured quasars)
we use the ratio of bolometric luminosity to monochromatic luminosity $L_{2500}$
(at 2500~\AA) from \cite{Elvis94}, along with the newly derived
optical---X-ray spectral index $\alpha_{\rm OX}$ given in
\cite{Elvis02}. We then use the correlation between X-ray luminosity and
radio luminosity given in Section~\ref{sec:xlf} to estimate the radio
luminosity at 200~MHz \citep[using $\alpha = 0.7$;][]{Kukula98}.

\begin{figure*}
{\hbox to \textwidth{\epsfxsize=0.4\textwidth
\epsfbox{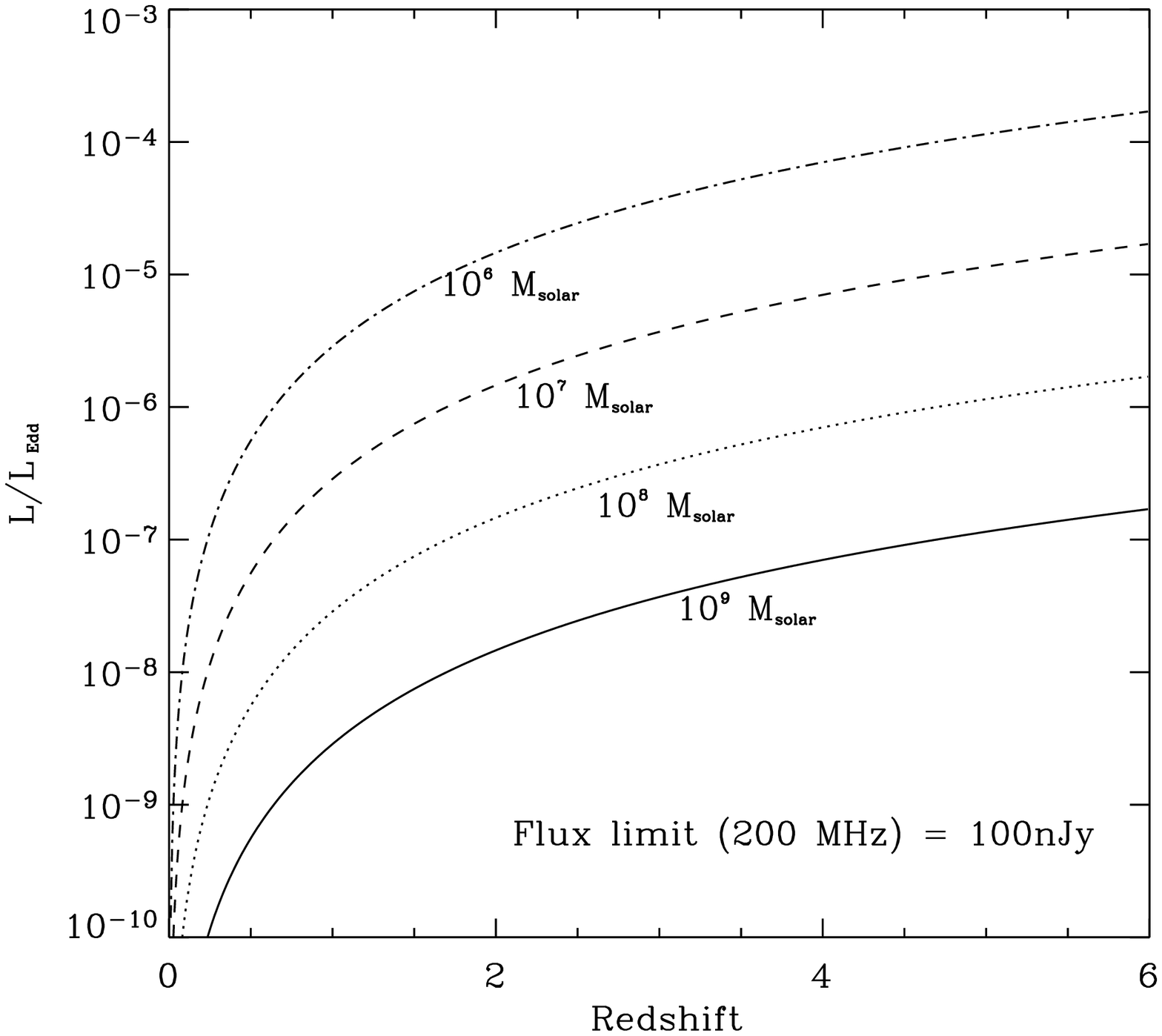}
\epsfxsize=0.4\textwidth
\epsfbox{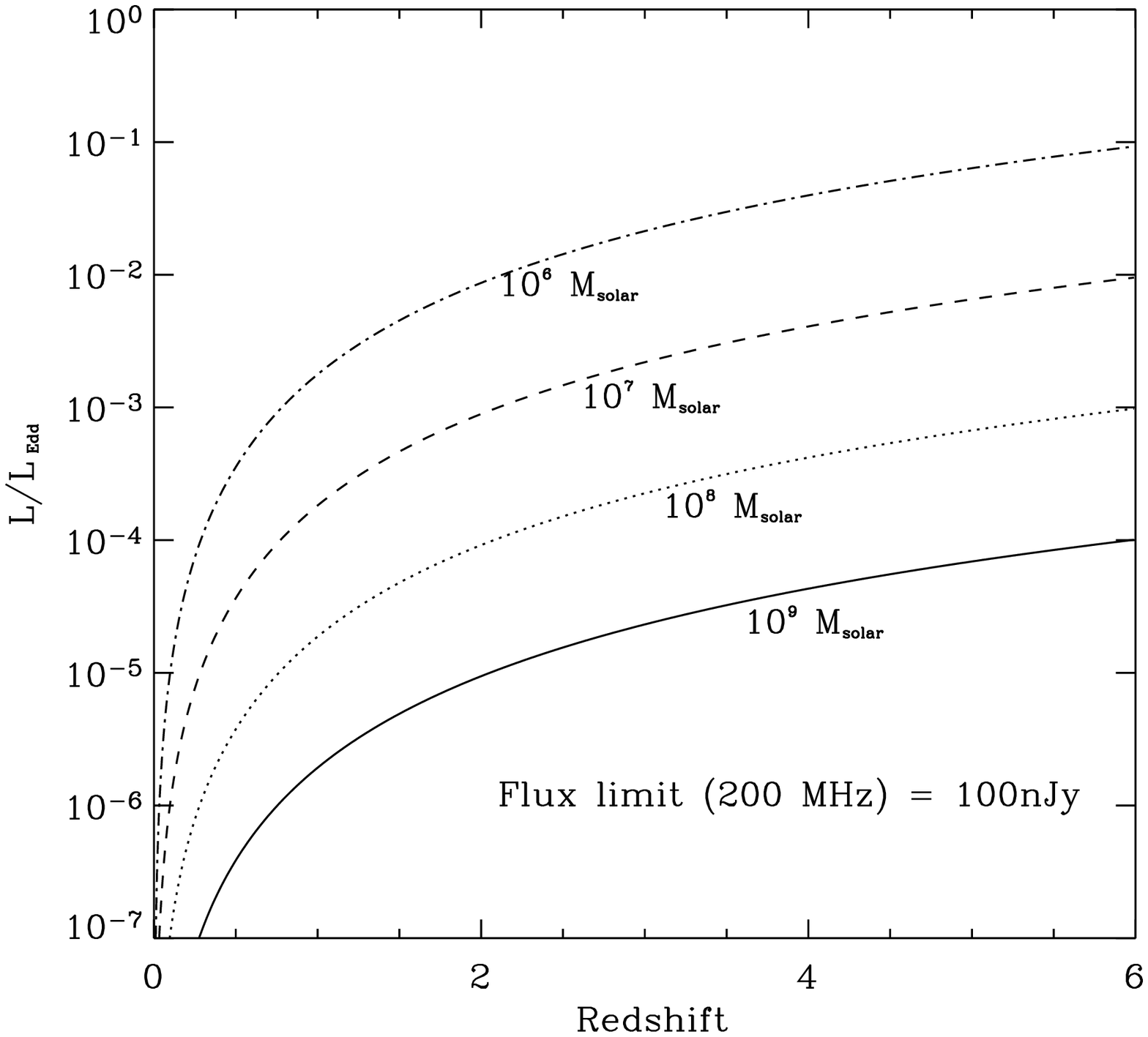} }}
{\caption{\label{fig:accretion} The detectable accretion luminosity
$L/L_{\rm Edd}$ versus redshift for sources containing black holes of
various mass ($10^{6} \rightarrow 10^{9}$~M$_{\odot}$), at a
flux-density limit of $S_{200\rm MHz} = 100$~nJy for radio-loud, typically FRII, 
sources (left) and for radio-quiet sources (right).}}
\end{figure*}

Fig.~\ref{fig:accretion} shows what the SKA, with a flux-density limit
of $S_{200 \rm MHz} > 100$~nJy, will see in terms of radio-loud
quasars (left) and radio-quiet quasars (right) as a fraction of the
Eddington luminosity versus redshift for AGN with black-hole masses
$10^{6} ~ {\rm M}_{\odot} < M_{\rm bh} < 10^{9}{\rm M}_{\odot}$. Given
that {\it powerful} radio sources seem to require a supermassive black
hole with mass $\gta 10^{8}$~M$_{\rm bh}$
\citep{Dunlop03,JarvisMcLure02,rhalfs1,mj2004} then it is obvious from
Fig.~\ref{fig:accretion} that the SKA will easily detect all such
sources. However, more interestingly the SKA will be able to detect all the 
radio-quiet quasars with $M_{\rm bh} > 10^{6}$~M$_{\odot}$ accreting at
$< 10$~per cent of its Eddington limit up to $z \sim 6$. In the case of the powerful 
quasars being found at high redshift in the SDSS, with black-hole masses 
$> 10^{9}~ {\rm M}_{\odot}$ \citep{CJWz6.4}, these would be detected up to 
$z \sim 6$ if their accretion luminosities were $\sim 10^{-4}~L_{\rm Edd}$, 
i.e. in the realm of what we now consider as dormant black holes.

\section{The optimal SKA design}\label{sec:skadesign}

In this section we investigate the most efficient design of the SKA
for determining the large scale properties and evolution of the
AGN. We only consider those AGN which have an X-ray luminosity
$L_{\rm X} > 10^{35} ~ \rm W$, i.e. those in the regime where accretion
onto a supermassive black hole is the only way of producing such
emission and avoiding luminosities where supernovae and X-ray binaries
may begin to dominate the population.
This means that the SKA will be able to probe the `whole AGN population'
up to $z \sim 6$ by just surveying down to micro-Jansky levels.

One of the crucial points in a survey to map the AGN population at $0
< z < 6$ with the SKA is that IT MUST be able to observe across the frequency
range $200 < \nu < 1400$~MHz. This
is not for the traditional reason of previous low-frequency-based redshift surveys
such as 6CE, 6C* and 7CRS. In these samples the low survey frequency selected
sources in an unbiased manner with respect to source orientation, 
i.e. sources are selected on their optically-thin lobe emission. Low-frequency selection 
is also used to target the highest
redshift radio galaxies \citep[see e.g.][]{Cohen04,Jea04}, whereas traditional high-frequency 
selection would miss
them due to the shape of radio galaxy spectra. The sensitivity of the SKA will mean that 
this type of `low-frequency
selection' is unimportant, as all of the radio sources will easily
be detected at 1.4~GHz regardless of the redshift.

However, the crucial advantage that the SKA will have over future
X-ray missions in probing the AGN population will be its ability to 
measure redshifts of all the gas-rich galaxies out to $z \sim 2$ 
\citep{Abdalla04} via HI
emission. This means that the redshifts of the majority of the source
counts at these flux-density levels will already be known. This paves
the way for both directly constraining the evolution in the high-redshift population along with their three-dimensional clustering, and to investigate the accretion rate of {\it all powerful} AGN from $z= 0 \rightarrow 2$.

Moreover, powerful AGN, invariably reside in massive galaxies
\citep[i.e. $> L^{\star}$;][]{Jarvis01b,7CKz,rhalfs1} and 
high-redshift-AGN activity, triggered by mergers, are likely to contain
higher gas fractions than at lower redshifts. Evidence for this has been seen in the past few years \citep[e.g.][]{Pap01,Greve04}

Therefore, the HI
emission signature will probably be detected to higher redshifts
than for the less massive galaxies which are detailed in \cite{Abdalla04}. 
There is also the possibility of utilizing the fact
that many AGN, both optically selected and radio selected, exhibit
some level of HI absorption in their Lyman-$\alpha$ profiles,
\citep[e.g.][]{vano97,Jarvis03,Wilman04}. This gas could also be
detected against the radio continuum of these distant
sources via the HI transition. However, this obviously needs the SKA to 
operate at lower and lower frequencies depending on the maximum redshift one wishes to
probe. 

In order to estimate whether the SKA will be able to determine
redshifts via HI absorption against radio-loud objects up to $z \sim
6$ (see also Morganti et al., this volume)
we assume a minimum HI column density of $N_{\rm HI} =
10^{23}$~m$^{-2}$ (or $N_{\rm HI} = 10^{19}$~cm$^{-2}$) 
and a line width of $\sim 100$~km~s$^{-1}$
which is similar to the absorption systems observed in
Lyman-$\alpha$ around high-redshift radio galaxies,
\citep{Jarvis03,Wilman04}.

Using equation 1 from \cite{Chengalur00} with a spin temperature of
$T_{\rm S} = 100$~K \citep[see e.g.][]{Beswick04} and unity covering factor we find a target
velocity-integrated optical depth $\tau {\rm d}V \approx 0.06 ~ \rm km
~ s^{-1}$.  Taking a source at $z \sim 6$ with a luminosity around the
break in the RLF for powerful radio sources, i.e. $L_{200} \sim
10^{26.5}$~W~Hz$^{-1}$~sr$^{-1}$ gives a flux density at the redshifted
frequency of HI of $S_{200\rm MHz} \approx 15$~mJy. In the proposed one-year all-sky survey, 
the expected dip due to HI absorption 
(at a level $\sim 6 \times 10^{-4}$ of the continuum) would be detected at
$\approx 10 \sigma$ in an optimally-smoothed spectrum.\footnotemark

\footnotetext{
Assuming $FOV = 1 ~ \rm deg^{2}$. If $FOV$ was (say) 10-times higher, these
features would be detected at $\approx 30 \sigma$ and weaker (lower $N_{\rm H}$ and
higher $T_{\rm S}$) features could be detected.
}

Obviously, ignoring any cosmic evolution in the 
HI properties, the situation\footnotemark becomes much better at lower redshift where 
higher sensitivities are reached for a source of the
same luminosity as the one considered at $z \sim 6$. Therefore, we envisage that the SKA
will be able to detect and measure the redshift of almost every
`FRII-like' AGN in the Universe up to at least $z \sim 6$, if these
sources lie within a shell of HI gas with column densities $\sim
10^{19}$~cm$^{-2}$, as has been observed around young powerful sources
at high redshift via resonant HI absorption within the Lyman-$\alpha$ emission line
\citep{Jarvis03,Wilman04}.
Redshifts of radio sources with lower HI column densities and/or higher spin temperatures
could also be measured if their intrinsic linewidths were narrower, or if their continuum
sources were brighter (see also Kanekar \& Briggs, this volume). 
With HI emission detectable at low redshift, and the hope that HI absorption
is close to ubiquitous at high redshift,
an almost complete census of the evolution of powerful radio AGN activity may well
be obtained using just the SKA.

\footnotetext{
To obtain full sky coverage for HI at all redshifts in the range
$0 < z < 6$ the continuum survey will need to adopt the `tiling' approach discussed
in Sec.\ 2 of Rawlings et al. (this volume).
}

\section{Conclusions}\label{sec:conclusions} 

In this paper we concentrate on $z < 6$ AGN (for a discussion of 
higher-redshift AGN, see Carilli et al., this volume). Although detecting radio emission
from $z < 6$ AGN  will be a trivial exercise for the SKA, measuring their 
redshifts and probing them in detail requires low-frequency capabilities. A SKA which can
operate between 200~MHz $\rightarrow 1400$~MHz would therefore be ideal for
investigating AGN activity after the epoch of reionization.
Here we highlight some of the science that the SKA could deliver 
regarding this population. We should be able to achieve the following.

\begin{itemize}

\item Detect all the $> 10^{6}$~M$_{\odot}$ black holes in the observable 
Universe with accretion luminosities above $\sim  0.1~L_{\rm Edd}$.

\item Trace accretion luminosities from $L/L_{\rm Edd} = 1 \rightarrow
  10^{-4}$ as a function of redshift up to $z = 6$ for all black holes
  with $M > 10^{9}$~M$_{\odot}$. The SKA will also be able to probe
  the $M > 10^{8}$~M$_{\odot}$ dormant ($L/L_{\rm Edd} < 10^{-3}$) black 
hole population out to $z \sim 6$.

\item Constrain the evolution in the radio luminosity function for
both radio-quiet and radio-loud AGN, up to $z \sim 2$, {\it without the
need for optical follow-up time}. This is possible because the SKA will
be able to detect HI emission from all massive galaxies up to this
redshift for a survey in which the effective exposure time is $\sim 4-20$ hours
(see Rawlings et al., this volume). 
Constraining the luminosity function at higher redshifts, i.e.
within the `quasar epoch' at $z \sim 2 \rightarrow 6$, requires (if
additional optical or near-infrared observations are to be avoided) HI absorption
experiments with an SKA that operates down to 200 MHz (see also
Kanekar \& Briggs, this volume).

\item Constrain the three-dimensional clustering of these sources,
  with respect to the general galaxy population (see van der Hulst et al., this
  volume), again without the
  need for optical follow-up time. This may be crucial to our
  understanding of how AGN activity is triggered with respect to the
  large-scale environment, i.e. where does the star formation occur in
  a cluster and where does the AGN activity occur. Does it occur along
  dark-matter filaments with AGN at their centres?  Progress will have been made on this
  issue by the time the SKA starts to operate, but only the SKA will 
  trace this evolution in an `all hemisphere' sample up to $z \sim 2$.

\item The huge cosmic volumes (in comparison to present and future
optical and X-ray telescopes) surveyed by the SKA would allow investigations of the
largest structures in the Universe. \cite{Brand03} have shown that
moderately powerful radio galaxies are able to trace structures on
$\sim 100$-Mpc scales. On discovering such structures the
SKA will be able to probe where the AGN and star-formation activity
occurs with respect to the large-scale structure. Counts of these
structures have cosmological implications (Blake et al., this volume;
Rawlings et al., this volume).  

\item The sensitivity of the SKA to the HI line will allow
extremely detailed investigations of the neutral gas surrounding the
central engine, especially with VLBI capability (Morganti et al., this volume). 
This high column density gas ($N_{\rm H} > 10^{26} ~ \rm m^{-2}$),
may be responsible for the archetypal type-II quasars and could
easily be detected with the SKA for any source with a moderately
bright core, up to the highest redshifts. Not only will the SKA be
able to detect it but it will also be able to trace the dynamics
\citep{Beswick04}. This sort of study will never be achieved with
optical telescopes because the high-column densities involved
lead to a heavily saturated absorption profile in Lyman-$\alpha$. 

\item Measure inflows/outflows of HI gas, both in emission and
absorption up to $z \sim 2$. 
The SKA would also have the power to see gas stripping by/around
an AGN by tracing the HI-emitting \citep[e.g.][]{Kenney04}. If powerful AGN lie at the
interface of merging sub-clusters \citep[e.g.][]{SimpsonSR02},
then the gas stripping around galaxies due to their
interaction with the intracluster medium should be easily detectable
with the SKA. Outflows of gas, driven by the AGN, may strip gas from not only the
host galaxy but also neighbouring galaxies.
This is one area where a multi-wavelength approach, particularly with
optical/near-infrared telescopes, would enable the investigation of
gas striping along with chemical enrichment within the intracluster
medium, and how this is affected by the geometry of any sub-cluster merger.

\item An `all-hemisphere' survey, deep enough to detect all the AGN in
the Universe, and measure redshifts for all of the most powerful ones, will
also provide the perfect sample for probing the magnetic fields along
the line of sight via Faraday Rotation measurements (Gaensler,
Beck \& Feretti, this volume). As the SKA will detect AGN at all
redshifts then any evolution in the magnetic fields could also be
investigated with such a survey.

\item The SKA would greatly benefit 
investigations of feedback mechanisms in the high-redshift Universe
where AGN-driven feedback events may be a crucial factor in the
evolution of all galaxies. In \cite{RJ04} we show that powerful radio
jets may have a profound influence on the evolution of all galaxies
surrounding the AGN. This is because the radio source is able to
inject enough energy into its surroundings that it gravitationally
unbinds ionized gas associated not only with its host galaxy, but more
widely throughout the protoclusters in which they 
seem to reside \citep{Venemans02}.
This process would yield a reservoir of gas within the protocluster
which is gravitationally unbound, and has been heated such that it
cannot accrete back onto the protogalaxies. Hence, there will be a
protocluster-wide shut down of activity, whether it be circumnuclear
star formation or black-hole accretion.
With the SKA we will be able to trace the neutral HI gas in such
protoclusters and trace the effect that the powerful radio jets have
on the protocluster environment. This will essentially be monitoring
AGN feedback mechanisms {\it in action}, as the SKA will have both the
sensitivity and the field-of-view to do this for {\it all} powerful AGN
at {\it all} redshifts.

\end{itemize}

\section*{ACKNOWLEDGEMENTS} 
MJJ acknowledges the support of a PPARC PDRA. SR is grateful to the UK 
PPARC for a 
Senior Research Fellowship, and for financial support from the Australia 
Telescope National Facility. We would also like to thank Richard Wilman 
for providing a suite 
of X-ray spectra of AGN and Nick Seymour for providing a compilation of 
radio source 
counts.

\end{document}